\documentclass[prd,superscriptaddress,unsortedaddress,twocolumn,showpacs,preprintnumbers,amsmath,amssymb]{revtex4}
\usepackage[dvips]{graphicx}
\usepackage{amsmath,amssymb,times}

\newcommand{\bequ}{\begin{equation}}
\newcommand{\eequ}{\end{equation}}
\newcommand{\bea}{\begin{eqnarray}}
\newcommand{\eea}{\end{eqnarray}}

\def\gsim{~\,\makebox(1,1){$\stackrel{>}{\widetilde{}}$}\,~}
\def\lsim{~\,\makebox(1,1){$\stackrel{<}{\widetilde{}}$}\,~}

\DeclareSymbolFont{boldletters}{OML}{cmm} {b}{it}
\DeclareSymbolFontAlphabet{\mathbit}{boldletters}
\DeclareMathSymbol{\alpha}{\mathalpha}{letters}{"0B}
\DeclareMathSymbol{\beta}{\mathalpha}{letters}{"0C}
\DeclareMathSymbol{\gamma}{\mathalpha}{letters}{"0D}
\DeclareMathSymbol{\delta}{\mathalpha}{letters}{"0E}
\DeclareMathSymbol{\epsilon}{\mathalpha}{letters}{"0F}
\DeclareMathSymbol{\zeta}{\mathalpha}{letters}{"10}
\DeclareMathSymbol{\eta}{\mathalpha}{letters}{"11}
\DeclareMathSymbol{\theta}{\mathalpha}{letters}{"12}
\DeclareMathSymbol{\iota}{\mathalpha}{letters}{"13}
\DeclareMathSymbol{\kappa}{\mathalpha}{letters}{"14}
\DeclareMathSymbol{\lambda}{\mathalpha}{letters}{"15}
\DeclareMathSymbol{\mu}{\mathalpha}{letters}{"16}
\DeclareMathSymbol{\nu}{\mathalpha}{letters}{"17}
\DeclareMathSymbol{\xi}{\mathalpha}{letters}{"18}
\DeclareMathSymbol{\pi}{\mathalpha}{letters}{"19}
\DeclareMathSymbol{\rho}{\mathalpha}{letters}{"1A}
\DeclareMathSymbol{\sigma}{\mathalpha}{letters}{"1B}
\DeclareMathSymbol{\tau}{\mathalpha}{letters}{"1C}
\DeclareMathSymbol{\upsilon}{\mathalpha}{letters}{"1D}
\DeclareMathSymbol{\phi}{\mathalpha}{letters}{"1E}
\DeclareMathSymbol{\chi}{\mathalpha}{letters}{"1F}
\DeclareMathSymbol{\psi}{\mathalpha}{letters}{"20}
\DeclareMathSymbol{\omega}{\mathalpha}{letters}{"21}
\DeclareMathSymbol{\varepsilon}{\mathalpha}{letters}{"22}
\DeclareMathSymbol{\vartheta}{\mathalpha}{letters}{"23}
\DeclareMathSymbol{\varpi}{\mathalpha}{letters}{"24}
\DeclareMathSymbol{\varrho}{\mathalpha}{letters}{"25}
\DeclareMathSymbol{\varsigma}{\mathalpha}{letters}{"26}
\DeclareMathSymbol{\varphi}{\mathalpha}{letters}{"27}
\DeclareMathSymbol{\Gamma}{\mathalpha}{letters}{"00}
\DeclareMathSymbol{\Delta}{\mathalpha}{letters}{"01}
\DeclareMathSymbol{\Theta}{\mathalpha}{letters}{"02}
\DeclareMathSymbol{\Lambda}{\mathalpha}{letters}{"03}
\DeclareMathSymbol{\Xi}{\mathalpha}{letters}{"04}
\DeclareMathSymbol{\Pi}{\mathalpha}{letters}{"05}
\DeclareMathSymbol{\Sigma}{\mathalpha}{letters}{"06}
\DeclareMathSymbol{\Upsilon}{\mathalpha}{letters}{"07}
\DeclareMathSymbol{\Phi}{\mathalpha}{letters}{"08}
\DeclareMathSymbol{\Psi}{\mathalpha}{letters}{"09}
\DeclareMathSymbol{\Omega}{\mathalpha}{letters}{"0A}

\begin{document}
\preprint{SAGA-HE-284, YITP-15-35}
\title{Understanding of QCD at high density from $Z_3$-symmetric
QCD-like theory}

\author{Hiroaki Kouno}
\email[]{kounoh@cc.saga-u.ac.jp}
\affiliation{Department of Physics, Saga University,
             Saga 840-8502, Japan}

\author{Kouji~~Kashiwa}
\email[]{kouji.kashiwa@yukawa.kyoto-u.ac.jp}
\affiliation{Yukawa Institute for Theoretical Physics, Kyoto University, Kyoto 606-8502, Japan}

\author{Junichi Takahashi}
\email[]{takahashi@phys.kyushu-u.ac.jp}
\affiliation{Department of Physics, Graduate School of Sciences, Kyushu University,
             Fukuoka 812-8581, Japan}

\author{Tatsuhiro Misumi}
\email[]{misumi@phys-h.keio.ac.jp}
\affiliation{Department of Physics, Keio University, Hiyoshi 4-1-1,Yokohama, Kanagawa 223-8521, Japan}
\affiliation{Department of Mathematical science, Akita University,
1-1 Tegata Gakuen-machi, Akita 010-8502, Japan}

\author{Masanobu Yahiro}
\email[]{yahiro@phys.kyushu-u.ac.jp}
\affiliation{Department of Physics, Graduate School of Sciences, Kyushu University,
             Fukuoka 812-8581, Japan}

\date{\today}

\begin{abstract}
We investigate QCD at large $\mu/T$ by using $Z_3$-symmetric $SU(3)$
 gauge theory, where $\mu$ is the quark-number chemical potential and
 $T$ is temperature.
We impose the flavor-dependent twist boundary condition on quarks in QCD. 
This QCD-like theory has the twist angle $\theta$ as a parameter, and
 agrees with QCD when $\theta=0$ and becomes symmetric when $\theta=2\pi/3$.
For both QCD and the $Z_3$-symmetric $SU(3)$ gauge theory, 
the phase diagram is drawn in $\mu$--$T$ plane
with the Polyakov-loop extended Nambu--Jona-Lasinio model.
In the $Z_3$-symmetric $SU(3)$ gauge theory, the Polyakov loop $\varphi$
 is zero in the confined phase appearing at $T \lsim 200$ MeV.
The perfectly confined phase never coexists with the color
 superconducting (CSC) phase, since finite diquark condensate in the CSC
 phase breaks $Z_3$ symmetry and then makes $\varphi$ finite.
When $\mu \gsim 300$ MeV, the CSC phase is more stable than the
 perfectly confined phase at $T \lsim 100$ MeV.
Meanwhile, the chiral symmetry can be broken in the perfectly confined
phase, since the chiral condensate is $Z_3$ invariant. 
Consequently, the perfectly confined phase is divided into the perfectly
 confined phase without chiral symmetry restoration in a region of $\mu
 \lsim 300$ MeV and $T \lsim 200$ MeV and the perfectly confined phase
 with chiral symmetry restoration in a region of $\mu \gsim 300$ MeV and
 $100 \lsim T \lsim 200$ MeV.
The basic phase structure of $Z_3$-symmetric QCD-like theory remains in QCD.
We show that in the perfectly confined phase the sign problem becomes
 less serious because of $\varphi=0$, using the heavy quark theory.
We discuss a lattice QCD framework to evaluate observables at $\theta=0$
 from those at $\theta=2\pi/3$.
\end{abstract}

\maketitle

\section{Introduction}
\label{sec:intro}

Quantum chromodynamics (QCD) has a lot of interesting phenomena
particularly at small temperature ($T$) and large quark-number chemical
potential ($\mu$). The phenomena may affect the structure of neutron
stars in its inner core.
The color superconducting 
 (CSC) phase that appears only at large $\mu/T$ is a typical example.
However, lattice QCD (LQCD) simulations as the first-principle
calculation have the well-known sign problem, since the fermion
determinant becomes complex for real $\mu$ and this makes the
importance-sampling method unfeasible in Monte Carlo
simulations. Therefore, several methods such as the reweighting method
\cite{Fodor}, the Taylor expansion method \cite{Allton,Ejiri_density}
and the analytic continuation from imaginary to real $\mu$
\cite{FP,D'Elia,D'Elia3,FP2010,Nagata,Takahashi} were proposed so far to
circumvent this problem, but these methods are considered to be reliable
only at $\mu/T \lsim 1$.
Recently, remarkable progress has been made with the complex Langevin 
dynamics~\cite{Aarts_CLE_1,Aarts_CLE_2,Sexty,Greensite,Aarts_CLE_3}
and the Lefschetz thimbles
approach~\cite{Aurora_thimbles,Fujii_thimbles}.
However, the results are still far from perfection.
Hence the effective model approach becomes useful. Actually, QCD
at large $\mu/T$ has been predicted with effective models such as
the Polyakov loop extended Nambu-Jona-Lasinio (PNJL)
model~\cite{Meisinger,Dumitru,Fukushima,Ratti,Megias,Rossner,
Schaefer,Abuki,Fukushima2,Kashiwa1,McLerran_largeNc,Sakai,Sasaki-T_Nf3}.

The LQCD and effective model approaches 
have a fundamental problem on the definition of quark confinement, as
mentioned below.
It is widely believed that the confinement mechanism in $SU(N)$ gauge
theory is governed by $Z_N$ symmetry and confinement and deconfinement
phases are defined by the Polyakov loop $\varphi$ \cite{Polyakov} as an
order parameter for $Z_N$ symmetry.
This is true for pure gauge theory, since $Z_N$ symmetry is exact.
The symmetry is preserved at small $T$, but spontaneously broken at high
$T$.
This makes the confinement-deconfinement transition first-order in the case of $N \ge 3$. 
In $SU(N)$ gauge theory with fundamental fermions, $Z_N$ symmetry is not
exact any more, so that the definition of confinement and deconfinement
phases becomes obscure.
In fact, in QCD with $N=3$, $\varphi$ is always finite at $T > 0$ and
the deconfinement transition is crossover for zero
$\mu$~\cite{YAoki_nature}.

An attempt to answer this problem was made recently by modifying the fermion boundary condition 
in Refs.~\cite{Kouno_TBC,Sakai_TBC,Kouno_adjoint,Kouno_TBC_2}.
Three degenerate flavor QCD was extended by imposing
the flavor-dependent twist boundary condition (FTBC)
\begin{eqnarray}
\Psi_f(\tau =\beta,{\bf x})=-e^{-i\theta_f}\Psi_f(\tau =0,{\bf x})
\label{FTBC_QCD}
\end{eqnarray}
on quarks, where $\beta=1/T$ and $\Psi_f$ is the quark field with flavor
$f$.
When the twist angles $\theta_f$ are set to
\bea
 \theta_1=\theta, \quad \theta_2=-\theta, \quad \theta_3=0,
\label{FTBC_angle}
\eea
the QCD-like theory agrees with QCD when $\theta=0$, and
becomes $Z_3$ symmetric when $\theta=2\pi/3$, since
$\theta_f$ is transformed into $\theta_{f-1}$ by
the $Z_3$ transformation but $\theta_{f-1}$ returns to $\theta_f$
by relabeling $f$. 
The $Z_3$-symmetric QCD-like theory with
$\theta=2\pi/3$ is referred to as $Z_3$-QCD in this paper.

Confinement and deconfinement phases are clearly defined with $\varphi$
in $Z_3$-QCD. The QCD phase diagram can be understood as a remnant of
the $Z_3$-QCD phase diagram. In $Z_3$-QCD, it is obvious that
there exists a confinement phase at small $T$ and a deconfinement phase
at high $T$, because $\varphi=0$ in the low-$T$ limit and 1
in the high-$T$ limit. 
We call the phase where $\varphi =0$ perfectly confined phase in this paper. 
The deconfinement phase transition is either first-order or second-order.
The order of the phase transition was investigated by applying
the FTBC to the PNJL model, and was found to be first-order
at $\mu=0$~\cite{Kouno_TBC,Sakai_TBC,Kouno_adjoint,Kouno_TBC_2}. 
LQCD simulations for 2+1 flavor quarks show that
$\varphi$ is quite small at zero $\mu$ and low $T$~\cite{BW_C_1,HQCD_1}.
The confinement property in 2+1 flavor QCD is considered
to be a remnant of the perfectly confined phase in $Z_3$-QCD.

$Z_3$-QCD is useful also to investigate the relation between
confinement and other mechanism such as chiral symmetry breaking and
CSC.
Particularly for the relation between confinement and CSC, we can make
clear discussion by using $Z_3$ symmetry.
The diquark condensate $\Delta$ as an order parameter for CSC is not
$Z_3$ invariant.
This means that $\Delta$ is an order parameter for both CSC and
$Z_3$ symmetry. The diquark condensate is then zero whenever
$\varphi$ is zero, and hence the CSC phase with finite
$\Delta$ never coexists with the perfectly confined phase 
with zero $\varphi$.
The correlation between $\varphi$ and the diquark condensate is thus
clearly understood in $Z_3$-QCD. 
The correlation in QCD can be understood as a remnant of that in $Z_3$-QCD.
Meanwhile, the chiral condensate as an order parameter for chiral
symmetry is $Z_3$ invariant. Hence, a chiral transition can take place
in the perfectly confined phase.
In fact, the PNJL model shows that this really happens at
finite $\mu$ \cite{Sakai_TBC}. In the PNJL analysis, however,
diquark effects were not considered.

In this paper, we try to understand QCD at high $\mu/T$ from $Z_3$-QCD.
The phase diagram is drawn in $\mu$--$T$ plane for both QCD and
$Z_3$-QCD by using the PNJL model with the FTBC. 
In the case of $Z_3$-QCD, a perfectly confined phase with $\varphi=0$
and $\Delta=0$ appears at $T \lsim 200$~MeV,
and the chiral restoration occurs at $\mu \approx 300$~MeV in the phase,
as expected. Hence, the perfectly confined phase is divided into
the perfectly confined phase without chiral symmetry restoration in
a region of $\mu \lsim 300$~MeV and $T \lsim 200$~MeV and
the perfectly confined phase with chiral symmetry restoration
in a region of $\mu \gsim 300$~MeV and $100 \lsim T \lsim 200$~MeV.
In a region of $\mu \gsim 300$~MeV and $T \lsim 100$~MeV,
CSC phases with finite $\Delta$ and small $\varphi$ come out, since they
cannot coexist with the perfectly confined phase and more stable than
the perfectly confined phase with chiral symmetry restoration.
This basic structure remains in QCD.
Second we show that in the perfectly confined phase the sign problem may
become less serious because of $\varphi=0$, using the heavy quark model.
This may make LQCD simulations feasible in the perfectly confined phase.
Particularly when the system is in the perfectly confined phase without
chiral symmetry restoration, the system changes continuously
from $Z_3$-QCD to QCD by varying $\theta$ from $2\pi/3$ to 0.
Using this property, we propose a way of estimating observables
for QCD from those for $Z_3$-QCD.

This paper is organized as follows.
We recapitulate the FTBC and $Z_3$-QCD in Sec. \ref{S_FTBC} and
the PNJL model in Sec. \ref{sec:PNJL}.
Numerical results are shown by using the PNJL model and
LQCD simulations in Sec. \ref{sec:NR}. The sign problem
in $Z_3$-QCD is discussed in Sec. \ref{sec:SP}.
Section \ref{Summary} is devoted to a summary.

\section{ $Z_3$-QCD }
\label{S_FTBC}

In this section, we recapitulate the FTBC and $Z_3$-QCD, following
Refs.~\cite{Kouno_TBC,Sakai_TBC,Kouno_adjoint,Kouno_TBC_2}.
We consider $SU(N)$ gauge theory with $N$ degenerate flavor quarks.
The Lagrangian density  ${\cal L}$ in Euclidean spacetime is
\bea
{\cal L}= \sum_{f=1}^{N} \bar{\Psi}_{f}(\gamma_\nu D_\nu + m) \Psi_{f}
+{1\over{4g^2}}{F_{\mu\nu}^{a}}^2,
\label{FD}
\eea
where $m$ is the current quark mass for all flavors and
$D_\nu \equiv \partial_\nu-iA_\nu$ for the gauge field $A_\mu$
and $F_{\mu\nu}$ is the field strength tensor.
The temporal boundary conditions for quarks are
\begin{eqnarray}
\Psi_{f}(\tau =\beta,{\bf x})=-\Psi_{f}(\tau =0,{\bf x}) .
\label{BC}
\end{eqnarray}
The Lagrangian density (\ref{FD}) is invariant under
the $Z_{N}$ (large gauge) transformation,
\begin{eqnarray}
\Psi_{f} &\to& \Psi_{f}^\prime =U \Psi_{f},
\nonumber
\\
A_\nu &\to& A_\nu^\prime =UA_{\nu}U^{-1}+i(\partial_\nu U)U^{-1},
\label{ZNtrans}
\end{eqnarray}
but the boundary condition (\ref{BC}) is changed into~\cite{RW}
\begin{eqnarray}
\Psi_{f}(\tau =\beta,{\bf x})=-e^{i2k\pi/N}\Psi_{f}(\tau =0,{\bf x}).
\label{BC_2}
\end{eqnarray}
Here
\begin{eqnarray}
U(x,\tau)&=&\exp{(i\alpha_at_a)}
\label{gauge_element}
\end{eqnarray}
is an element of $SU(N)$ group characterized by real functions
$\alpha_a(x,\tau)$ satisfying the boundary condition
\begin{eqnarray}
U (x,\beta )&=&\exp{(-i2\pi k/N)}U(x,0)
\label{ZN}
\end{eqnarray}
for any integer $k$.

Now, we consider the following FTBC instead of (\ref{BC}):
\begin{eqnarray}
\Psi_f(\tau =\beta,{\bf x})=-e^{-i\theta_f}\Psi_f(\tau =0,{\bf x}),
\label{FTBC_B}
\end{eqnarray}
where
\begin{eqnarray}
\theta_f=\theta_1+{2\pi\over{N_f}}(f-1)~~~~(f=1,2,\dots,N_f),
\label{thetaf}
\end{eqnarray}
with $N_f=N$.
Under the $Z_{N}$ transformation (\ref{ZNtrans}),
the boundary condition (\ref{FTBC_B}) is changed into
\begin{eqnarray}
\Psi_f(\tau =\beta,{\bf x})=-e^{-i\theta_f^\prime}\Psi_f(\tau =0,{\bf x}),
\label{FTBC_B_2}
\end{eqnarray}
where
\begin{eqnarray}
\theta_f^\prime =\theta_1+{2\pi\over{N}}(f-1-k)~~~~(f=1,2,\dots,N).
\label{thetaf_2}
\end{eqnarray}
The boundary condition \eqref{thetaf_2} returns to the original one
\eqref{thetaf} by relabeling the flavor index $f-k$ as $f$.
Hence $SU(N)$ gauge theory with the FTBC \eqref{FTBC_B}
has $Z_N$ symmetry exactly.
$SU(N)$ gauge theory with $l \times N$ flavor fundamental quarks also has
$Z_N$ symmetry for any positive integer $l$, when the FTBC \eqref{FTBC_B}
with $N_f=lN$ is imposed on the fermions~\cite{Kouno_TBC}.

When the fermion fields $\Psi_f$ are transformed as~\cite{RW}
\begin{eqnarray}
\Psi_f \to \exp{(-i\theta_fT\tau )}\Psi_f ,
\label{transform_1}
\end{eqnarray}
the boundary condition (\ref{FTBC_B}) returns to the ordinary one
(\ref{BC}), but the Lagrangian density ${\cal L}$ is changed into
\bea
{\cal L}^{\theta}= \sum_{f=1}^{N}
\bar{\Psi}_{f}(\gamma_\nu D_\nu^\theta + m){\Psi}_{f}
+{1\over{4g^2}}{F_{\mu\nu}^{a}}^2,
\label{FD_FTBC}
\eea
where $D_\nu^\theta \equiv \partial_\nu-i(A_\nu +\hat{\theta}
\delta_{\nu,4}T )$ and $i\hat{\theta}T$ is the flavor-dependent
imaginary chemical potential given by the matrix
\begin{eqnarray}
\hat{\theta} &=& {\rm diag}(\theta_1,\theta_2,\cdots,\theta_N)
\nonumber\\
&=& {\rm diag}(\theta_1,\theta_1 + 2\pi/N,\cdots,
\nonumber\\
&& \theta_1+2(f-1)\pi/N,\cdots,\theta_1+2\pi (N-1)/N)
\label{FTBC}
\end{eqnarray}
in flavor space.
In the case of $N=3$, ${\cal L}^{\theta}$ is nothing but the Lagrangian
density of $Z_3$-QCD.
The flavor-dependent imaginary chemical potential breaks $SU(N)$ flavor
symmetry and associated $SU(N)$ chiral symmetry
\cite{Kouno_TBC,Sakai_TBC}.
In the chiral limit, global $SU_{\rm V}(3)\times SU_{\rm A}(3)$ symmetry
is broken down to  $(U(1)_{\rm V})^2\otimes (U(1)_{\rm A})^2$
\cite{Kouno_adjoint}.
The symmetry is even broken into $(U(1)_{\rm V})^2$, when chiral
symmetry is spontaneously violated.

\section{PNJL model with flavor-dependent twist boundary condition }
\label{sec:PNJL}

In this section, we explain the Polyakov-loop extended
Nambu-Jona-Lasinio (PNJL) model with the FTBC \eqref{FTBC_B}
and diquark effects. 
Taking the Polyakov-gauge and treating $A_4$ as a background gauge
field, one can get the model Lagrangian density for three degenerate
flavors as
\bea
{\cal L}_{\rm PNJL}^\theta &=& \sum_{f=u,d,s}\bar{\Psi}_f(\gamma_\nu
{D}_\nu^{\theta} + m)\Psi_f+{\cal L}_{\rm NJL}^{\rm int} + {\cal U},
\label{PNJL}
\eea
where ${D}_\nu^{\theta}=\partial_\nu -i\delta_{\nu,4}(A_4+\hat{\theta}T)$.
In (\ref{PNJL}), ${\cal U}$ is a function of $\varphi$ and
its complex conjugate $\varphi^{*}$ defined by
\begin{eqnarray}
\varphi &=& {1\over{3}}{\rm tr}_c [e^{i A_4/T}]
\nonumber \\
&=&{1\over{3}}(e^{i\phi_1}+e^{i\phi_2}+e^{i\phi_3}), 
\end{eqnarray}
with the condition $\phi_1+\phi_2+\phi_3=0$.

We take the following form \cite{Rossner} as ${\cal U}$:
\begin{align}
&{\cal U} = T^4 \Bigl[-\frac{a(T)}{2} {\varphi}^*\varphi\notag\\
      &~~~~~+ b(T)\ln(1 - 6{\varphi\varphi^*}  + 4(\varphi^3+{\varphi^*}^3)
            - 3(\varphi\varphi^*)^2 )\Bigr] ,
            \label{eq:E13}\\
&a(T)   = a_0 + a_1\Bigl(\frac{T_0}{T}\Bigr)
                 + a_2\Bigl(\frac{T_0}{T}\Bigr)^2,~~~~
b(T)=b_3\Bigl(\frac{T_0}{T}\Bigr)^3 .
            \label{eq:E14}
\end{align}
The parameters in $\mathcal{U}$ are determined from
LQCD data~\cite{Boyd,Kaczmarek} in the pure gauge (heavy quark) limit.
The Polyakov-loop potential has a
first-order deconfinement phase transition at $T=T_0$ in the limit, and
hence $T_0=270$ MeV.
In the case of finite quark mass, however, the PNJL model
with this value overestimates the pseudocritical temperature
$T_c\approx 160$~MeV at zero $\mu$ determined by
full LQCD \cite{Borsanyi,Soeldner,Kanaya}.
We have then rescaled $T_0$ to 195~MeV to reproduce
$T_c\sim 160$~MeV~\cite{Sasaki-T_Nf3}.
The parameters in the Polyakov-loop potential are summarized in Table~\ref{table-para_Nc3}(a).

\begin{table}[h]
\begin{center}
\begin{tabular}{c|ccccc}
\hline \hline
\raisebox{-1.5ex}[0cm][0cm]{~(a)~}&$a_0$&$a_1$&$a_2$&$b_3$&$T_0$(MeV)
\\
\cline{2-6}
& 3.51 & -2.47 & 15.2 & -1.75 & 195
\\
\hline \hline
\raisebox{-1.5ex}[0cm][0cm]{(b)}&~$m_f$(MeV)~&~$\Lambda$(MeV)~&~$~G_{\rm s}\Lambda^2~$~&~$~G_{\rm D}\Lambda^2~$~& $K\Lambda^5$
\\
\cline{2-6}
&5.5&602.3&1.835& ${3\over{4}}G_{\rm S}\Lambda^2$ & 12.36
\\
\hline \hline
\end{tabular}
\caption{
Summary of the parameter set in the PNJL model for the case of $N=3$.
Panels (a) and (b) show the parameters in the Polyakov-loop potential and the NJL sector, respectively. }
\label{table-para_Nc3}
\end{center}
\end{table}

In (\ref{PNJL}), ${\cal L}_{\rm NJL}^{\rm int}$ stands for the effective quark-antiquark and quark-quark interactions 
~\cite{Ruster}:
\begin{eqnarray}
{\cal{L}_{\rm NJL}^{\rm int}}
&=&-G_{\rm S}\sum_{a=0}^{8}[({\bar \Psi}\lambda_a \Psi)^2+({\bar \Psi}i\gamma_5\lambda_a \Psi )^2]
\notag\\
&&-G_{\rm D}\sum_{\alpha =u,d,s}^4\sum_{c=r,g,b}[({\bar{\Psi}})_{\alpha}^{a} i\gamma_5\epsilon^{\alpha\beta\gamma}\epsilon_{abc}(\Psi_C)_\beta^b]
\notag\\
&&\times [{ (\bar{\Psi}_C) }_{\rho}^u i\gamma_5\epsilon^{\rho\sigma\gamma}\epsilon_{uvc}(\Psi )_\sigma^v]
\nonumber\\
&&+K\left[\det_{ff^\prime}{\bar \Psi}_f(1+\gamma_5)\Psi_{f^\prime} +{\rm h.c.}\right], 
\label{NJL-F}
\end{eqnarray}
where the $\lambda_a$ are the Gell-Mann matrices in flavor space,
$G_{\rm S}$ and $G_{D}$ are coupling constants of the four-quark
interactions and $K$ is a coupling constant of the
Kobayashi-Maskawa-'t Hooft determinant interaction~\cite{KMK,tHooft}. The
quark field $\Psi_\alpha^a$ carries color~($a=r,g,b$) and flavor
($\alpha =u,d,s$) indices.
The values of coupling constants, current quark mass and
three dimensional momentum cutoff $\Lambda$ are tabulated
in Table \ref{table-para_Nc3}(b).
The coupling constants and the cutoff were determined to reproduce
empirical values of $\eta'$- and $\pi$-meson masses and $\pi$-meson
decay constant at vacuum when $m_u=m_d=5.5$MeV and
$m_s=140.7$MeV~\cite{Rehberg}.
In this paper, however, we consider a symmetric current quark mass of
$m=m_f=5.5$ MeV to preserve SU($3$) flavor symmetry.

Taking the mean field approximation, one can get the thermodynamic
potential per volume as
\begin{eqnarray}
\Omega
&=&\Omega_q+U+{\cal U}
\nonumber\\
&&-\sum_{j=1 }^{2NN_f} \int \frac{d^3 p}{(2\pi)^3}
   \Bigl[{E}_{j}+{2\over{\beta}}
   \ln{ ( 1+e^{-\beta {E}_{j}} ) }
   \Bigr],
\label{Omega}
\end{eqnarray}
where the mean-field potential part $U$ is given by
\begin{eqnarray}
U&=&G_{\rm D}\sum_{l=1,2,3}|\tilde{\Delta}_l|^2+2G_{\rm S}\sum_{f=u,d,s}\sigma_f^2
\nonumber\\
&&-4K\sigma_u\sigma_d\sigma_s
\label{U_nc3}
\end{eqnarray}
with chiral condensates $\sigma_f=\langle \bar{\Psi}_f\Psi_f \rangle$
for $f=u,d,s$ and diquark condensates $\tilde{\Delta}_l=\langle
(\bar{\Psi}_C)_\alpha^ai\gamma_5\epsilon^{\alpha\beta
l}\epsilon_{abl}\Psi_\beta^b \rangle$ where no sum is taken over $l$
on the right hand side.
In (\ref{U_nc3}), the ${E}_{j}$ are the quark spectra which depend
on the absolute value of three-dimensional quark momentum ${\bf
p}$, the effective quark mass
\begin{eqnarray}
M_f=m_f-4G_{\rm S}\sigma_f+2K\sigma_{f^\prime}\sigma_{f^{\prime\prime}} ,
\nonumber\\
~~~(f\neq f^\prime,~f\neq f^{\prime\prime},~f^\prime \neq f^{\prime\prime})
\label{M_nc3}
\end{eqnarray}
the diquark condensate (multiplied by $-2G_{\rm D}$ )
\begin{eqnarray}
\Delta_l=-2G_{\rm D}\tilde{\Delta}_l ,
\label{Delta_nc3}
\end{eqnarray}
and the effective chemical potential
\begin{eqnarray}
\mu_f^c=\mu +i\phi_c T +i\theta_f T.
\label{mueff}
\end{eqnarray}
It is difficult to obtain the explicit forms of $E_j$ analytically, but
we can determine the $E_j$  by solving the eigenvalue equation for Dirac
operator numerically.
Using these spectra, we can calculate $\Omega$ and then find the
solutions $\sigma_f$, $\Delta_l$ and $\varphi$ that minimize $\Omega$.
Obviously, $\sigma_f$ is invariant under the $Z_3$ transformation
\eqref{ZNtrans}, but $\Delta_l$ is not.

\section{Numerical Results}
\label{sec:NR}

In this section, we show numerical results of the PNJL model.
For later convenience, we put
\begin{eqnarray}
\theta_u= \theta_1= \theta, \quad \theta_d=\theta_2=-\theta, 
\quad \theta_s=\theta_3=0 .
\label{theta_para}
\end{eqnarray}
The fermion boundary condition \eqref{theta_para} agrees with
the FTBC \eqref{thetaf} when $\theta ={2\pi/3}$ and the standard
antiperiodic boundary condition when $\theta =0$.

In $Z_3$-QCD with $\theta ={2\pi/3}$,
the flavor symmetry is violated with the FTBC 
(as the mentioned above about the flavor-dependent imaginary chemical potential),
but $u$ and $d$ quarks are symmetric under the interchange
between $u$ and $d$.
Since the boundary condition is changed by the $Z_3$ gauge
transformation, flavor and diquark indices are renamed so that
the conditions $M_1\leq M_2 \leq M_3$ and $|\Delta_1|\le |\Delta_2 |\le
|\Delta_3 |$ can be satisfied.

\begin{figure}[htbp]
\begin{center}
\includegraphics[width=0.5\textwidth]{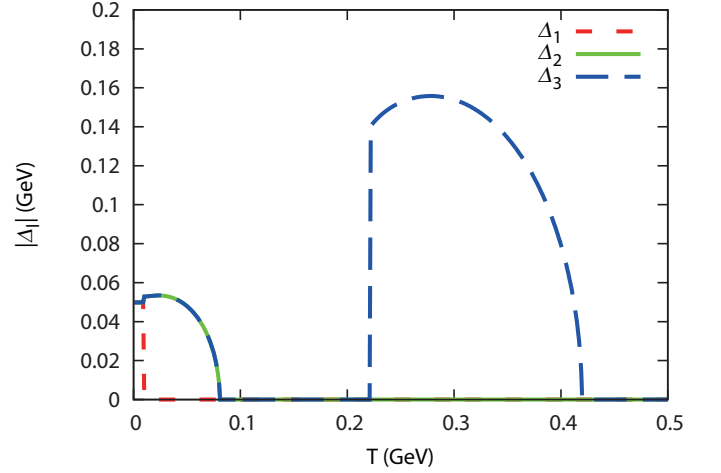}
\end{center}
\caption{$T$ dependence of $|\Delta_l |$
at $\mu =340$ MeV  in the PNJL model with $\theta =2\pi /3$. 
The diquark condensates $\Delta_1$, $\Delta_2$ and $\Delta_3$ are denoted
by dashed, solid and long dashed lines, respectively.
Below $T=10$ MeV, three diquark condensates exist.
In a region of $T=10\sim 80$ MeV, $\Delta_1=0$ and $\Delta_2=\Delta_3$.
}
\label{delta_s_f}
\end{figure}

Figure~\ref{delta_s_f} shows $T$ dependence of the absolute value of diquark condensates
$\Delta_1$, $\Delta_2$ and $\Delta_3$ at $\mu =340$ MeV for $Z_3$-QCD
with $\theta ={2\pi/{3}}$.
There appear a variety of CSC phases~\cite{Ruster}.
Below $T=T_1=$10 MeV, all the diquark condensates are finite,
indicating that it is the color flavor locking (CFL) phase.
Two of three condensates are finite in the region $T_1\le T\le T_2=80$ MeV,
and one of three is finite in the region $T_3=223\le T\le T_4=419$ MeV.
This means that the former is the uSC phase and the latter is the 2SC phase.

\begin{figure}[htbp]
\begin{center}
\includegraphics[width=0.5\textwidth]{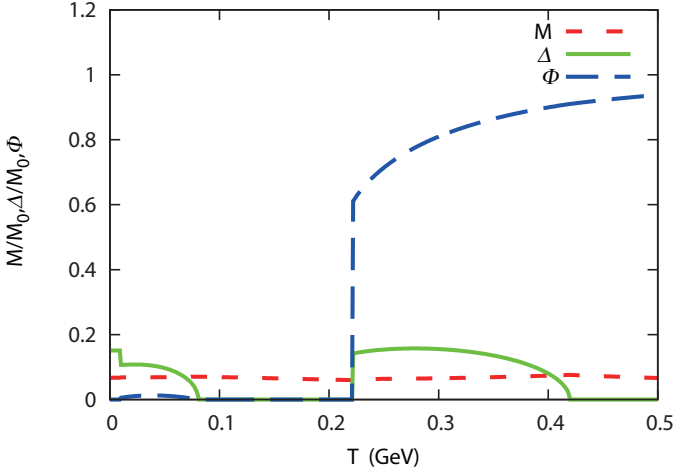}
\end{center}
\caption{$T$ dependence of $M$, 
 $\Delta$, and $\Phi$ at $\mu =340$ MeV in
 the PNJL model with $\theta = 2\pi /3$.
Here, $M$, $\Delta$ and $\Phi$ are denoted by dashed, solid and long
 dashed lines, respectively, and $M$ and $\Delta$ are normalized by the
 constituent quark mass $M_0$ at the vacuum.
}
\label{order_s_f}
\end{figure}

In Fig.~\ref{order_s_f}, the average values of dynamical quark masses
and the absolute value of diquark condensates, namely, $M=(M_1+M_2+M_3)/3$ and
$\Delta=(|\Delta_1 |+| \Delta_2 |+|\Delta_3 |)/3$, are plotted as a function of
$T$, together with $\Phi \equiv |\varphi |$, at $\mu =340$ MeV for
$Z_3$-QCD with $\theta ={2\pi/{3}}$.
There are three equivalent solutions to $\varphi$:
$\varphi = \Phi e^{i\phi}$ with $\phi =0,{\pm 2\pi /3}$ ( or $\phi =\pi,
{\pm \pi/3}$).
The solutions of $\phi ={\pm 2\pi /3}~({\pm \pi /3})$ are $Z_3$ images
of the solution of $\phi =0~(\pi)$.
In fact, they are thermodynamically equivalent and are transformed from
one to another by the $Z_3$ transformation.
We can then take the solution of $\phi =0~(\pi)$, and $\varphi$ is
real.
As mentioned in Sec. \ref{sec:intro}, finite diquark condensates in CSC
phases break $Z_3$ symmetry and hence induce finite $\Phi$.
In the CFL and uSC phases below $T_2$, $\Phi$ is tiny but finite, as
expected.
Thus, CSC phases do not coexist with the perfectly confined phase,
but with the almost-confined phase.
Also in the 2SC phase at $T_3<T<T_4$, $\Phi$ is finite as expected, but
it is large. Thus quarks are deconfined in the 2SC phase.
In the region $T_2<T<T_3$, $\Phi$ is zero and hence diquark condensates
are also zero. This is the perfectly confined phase.
Above $T_4$, $\Phi$ is finite but $\Delta$ is zero.  This is the pure
deconfinement phase without diquark condensate.

\begin{figure}[htbp]
\begin{center}
\includegraphics[width=0.5\textwidth]{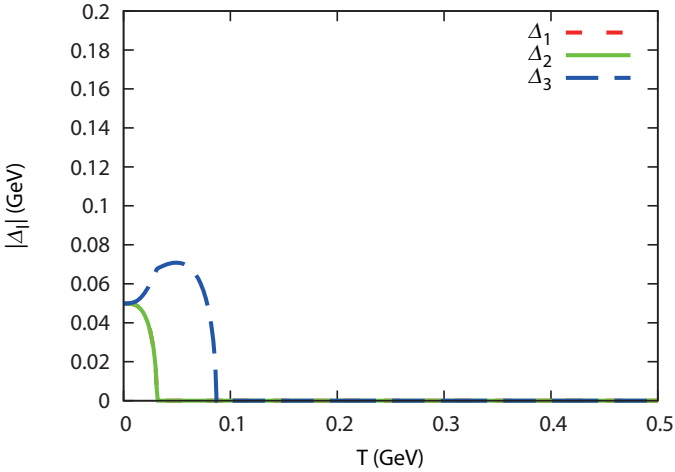}
\end{center}
\caption{$T$ dependence of $|\Delta_l |$ at $\mu
 = 340$ MeV in the PNJL model with $\theta =0$. 
See Fig. \ref{delta_s_f} for the definition of lines.
}
\label{delta_s_u}
\end{figure}

\begin{figure}[htbp]
\begin{center}
\includegraphics[width=0.5\textwidth]{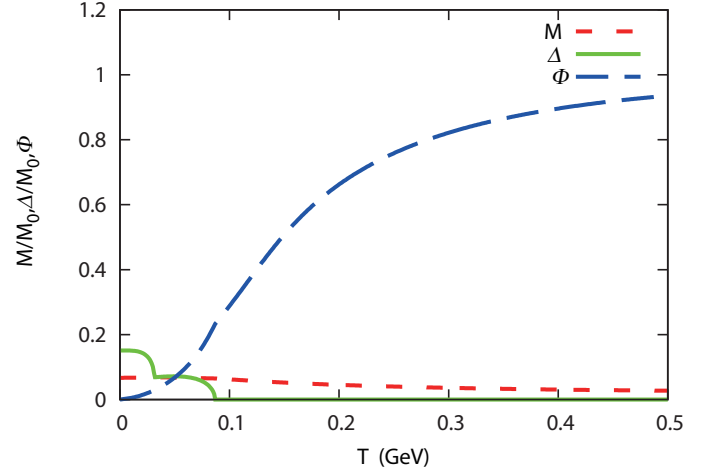}
\end{center}
\caption{$T$ dependence of $M$, $\Delta$ and $\Phi$ at $\mu =340$ MeV
 in the PNJL model with $\theta=0$. Here $M$ and $\Delta$ are divided by
 $M_0$.
See Fig. \ref{order_s_f} for the definition of lines.
}
\label{order_s_u}
\end{figure}

Figure~\ref{delta_s_u} is the same as Fig. \ref{delta_s_f}, but for QCD
with $\theta=0$.
Below $T=T_1=$32 MeV, all $\Delta_l$ are finite, indicating that it is
the CFL phase.
In the region $T_1\le T\le T_2=86$ MeV, one of three appears.
It is the 2SC phase. The phase structure is thus much simpler at $\theta
=0$ than at $\theta =2\pi/3$.
Figure~\ref{order_s_u} is the same as Fig. \ref{order_s_f}, but for
$\theta =0$.
In this case, $\varphi$ is always real and $\varphi =|\varphi |=\Phi$.
Since $Z_3$ symmetry is not exact in this case, $\Phi$ is always finite
at $T > 0$.
Thus we cannot define the confinement and deconfinement phases clearly.
Since $\Phi$ is small in the CFL and 2SC phases, one can consider that
the CSC phases are in the almost-confined phase.

\begin{figure}[htbp]
\begin{center}
\includegraphics[width=0.5\textwidth]{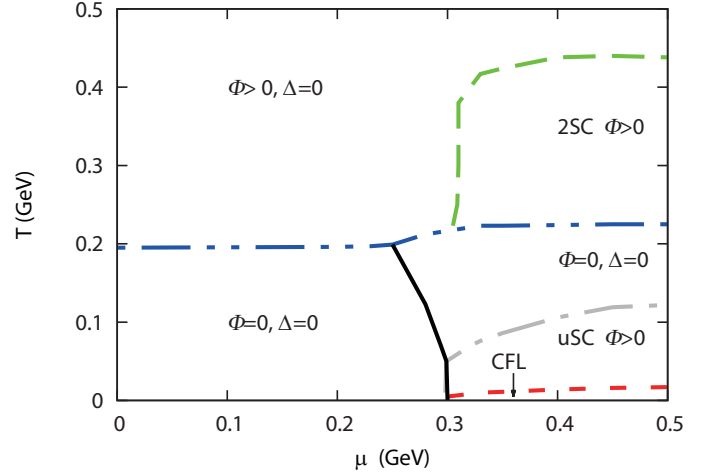}
\end{center}
\caption{Phase diagram in the PNJL model with $\theta =2\pi /3$. 
The CFL phase exists below the dotted line.
The solid line stands for the first-order chiral phase transition; $M$
 is large on the left side of the solid line, but small on the right
 side.
The dashed double-dotted line denotes the first-order deconfinement
 phase transition; the system is in the confinement phase below the line
 and in the deconfinement phase above the line.
The perfectly confined phase is labeled by ``$\Phi=0,~\Delta=0$", while
 the CFL, uSC and 2SC phases are by ``CFL",``uSC" and ``2SC", respectively.
}
\label{phase_s_f}
\end{figure}

Figure~\ref{phase_s_f} shows the phase diagram in $\mu$--$T$ plane
for the case of $\theta ={2\pi/3}$.
A variety of phases are present in this case.
When $T$ increases from zero with $\mu$ fixed at a large value,
say $350$~MeV, the CFL, uSC, perfectly-confined, 2SC and pure deconfinement
phases appear.
We also see that the chiral transition takes place in
the perfectly confined phase.

\begin{figure}[htbp]
\begin{center}
\includegraphics[width=0.5\textwidth]{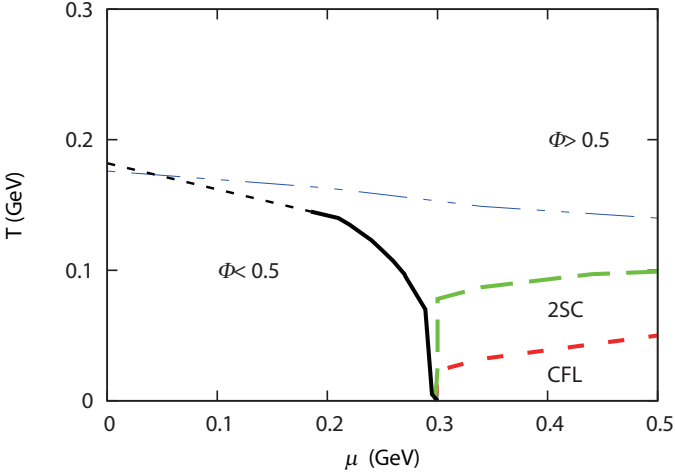}
\end{center}
\caption{Phase diagram in the PNJL model with $\theta =0$. 
The solid line (thin dotted line) stands for the first-order (crossover)
 chiral transition.
The thin dashed double-dotted line denotes the line of $\Phi =0.5$.
The almost-confined phase is labeled by ``$\Phi<0.5$", while the
 deconfined phase is by ``$\Phi > 0.5$" and the CFL and 2SC phases are
 by ``CFL" and ``2SC", respectively.
}
\label{phase_s_u}
\end{figure}

Figure~\ref{phase_s_u} is the same as Figure~\ref{phase_s_f} but for
$\theta =0$.
The uSC  phase disappears and the 2SC goes down to lower $T$, while
at small $T$ the CFL phase remains and the perfectly-confined phase becomes
an almost-confined phase without diquark condensate.
The almost-confined phase and the CFL phase at low $T$ can thus be
regarded as remnants of the perfectly confined phase and the CFL phase
in $Z_3$-QCD.

\section{Sign Problem in $Z_3$-QCD}
\label{sec:SP}

It is well known that LQCD has the sign problem at real $\mu$ and it
makes the importance sampling method unfeasible.
Also in the case of $Z_3$-QCD, the sign problem exists in principle.
In fact, the determinant of Dirac operator ${\cal M}$ has the following
relation~\cite{Sakai_TBC}:
\begin{eqnarray}
\left[\det{\cal M}(\mu_f)\right]^*&=&
\det{\cal M}(-\mu_f^*)\notag\\
&=&\prod_{f=u,d,s}\det[D-(\mu-i\theta_fT)\gamma_4+m_f]
\notag\\
&=&\prod_{f=u,d,s}\det[D-(\mu+i\theta_fT)\gamma_4+m_f]
\notag\\
&=&\det{\cal M}(-\mu_f)
\label{E_fermion_det}
\end{eqnarray} 
for $\mu_f=\mu +i\theta_fT$ satisfying the condition \eqref{theta_para},
where the third equality has been obtained by relabeling the $f$.
Hence, the determinant ${\rm det}{\cal M}(\mu_f)$ is not real.
The present system thus has the sign problem, although the partition
function is real, since
\bea
Z(\mu_f)^*=Z(-\mu_f)=Z(\mu_f),
\label{E_Omega_reality}
\eea
where the first equality comes from \eqref{E_fermion_det} and the second
one from charge conjugation.
Note that Eq. (\ref{E_Omega_reality}) is true for any value of $\theta$.

Although $Z_3$-QCD has the sign problem at real $\mu$, there is a
possibility that it is not serious in the perfectly-confined
phase. 
To see this clearly, we consider the heavy quark
model~\cite{Greensite,Aarts_CLE_3}.
In the model, the fermion determinant of Dirac operator is given
in terms of the Polyakov line operator ${\rm Tr}[U_{\rm x}]$:
\bea
\det{\cal M}(\mu_f)={\rm det}[1+he^{\mu/T}U_{\rm x}]{\rm det}[1+he^{-\mu/T}U_{\rm x}^\dagger]
\eea
with
\bea
&&{\rm det}[1+he^{\mu/T}U_{\rm x}]=1+he^{\mu /T}{\rm Tr}[U_{\rm x}]
\nonumber\\
&& \hspace{3cm} +h^2e^{2\mu /T}{\rm Tr}[U_{\rm x}^\dagger]+h^3e^{3\mu/T},
\nonumber\\
&&{\rm det}[1+he^{-\mu/T}U_{\rm x}^\dagger]=1+he^{-\mu /T}{\rm Tr}[U_{\rm x}^\dagger]
\nonumber\\
&& \hspace{3cm} +h^2e^{-2\mu /T}{\rm Tr}[U_{\rm x}]+h^3e^{-3\mu/T},
\nonumber\\
\label{det_h_q}
\eea
where $U_{\rm x}=U({\rm x},0)$ for the lattice link variable $U({\rm x},0)$
depending on the space coordinate ${\bf x}$ only.
The parameter $h$ is
defined as $h=\zeta^{N_t}$ with the lattice-site number $N_t$
in the temporal direction and the hopping parameter $\zeta$
for Wilson fermion and ${1/({2m})}$
for staggered fermion with a current quark mass $m$.
In general, $\det{\cal M}(\mu_f)$ is not real, since the exponential
factor with ${\rm Tr}[U_{\rm x}]$ is not equal to that with its
conjugate ${\rm Tr}[U_{\rm x}]^\dagger$.
For the configurations that satisfy the condition ${\rm Tr}[U_{\rm x}]=0$,
however, only the first and fourth terms remain on the right hand side of
of (\ref{det_h_q}), and hence  $\det{\cal M}(\mu_f)$ becomes real.
In $Z_3$-QCD, we use $\mu_f$ instead of $\mu$, but the factor $e^{\pm
3\mu_f/T}=e^{\pm 3\mu}$ is still real when $\theta =2\pi /3$.
This indicates the possibility that the sign problem may not be serious
in the perfectly confined phase, since the gauge configurations are
concentrated on the vicinity of $\Phi=0$.
One may then perform the importance sampling in lattice $Z_3$-QCD. 

Once a thermodynamical quantity $O(\theta)$ and it derivatives
with respect to $\theta$ are obtained at $\theta ={2\pi/{3}}$, the
quantities at $\theta$ less than ${2\pi/{3}}$ are obtainable
by using the Taylor expansion
\begin{eqnarray}
O(\theta)
=\sum_{n=0}^\infty
 {1\over{n!}}
 \left.{\partial^n O(\theta )\over
       {\partial \theta^n}}\right|_{\theta =2\pi/3}
 \Bigl( \theta -{2\pi\over{3}} \Bigr)^n .
\label{Taylor}
\end{eqnarray}

First we consider the perfectly confined phase with chiral symmetry
breaking in $Z_3$-QCD. The phase is located in a region of $T \lsim
200$~MeV and $\mu \lsim 300$~MeV. In QCD with $\theta=0$, it is very
likely that the region includes the nuclear-matter region at $\mu
\approx 300$ MeV and $T=0$.

\begin{figure}[htbp]
\begin{center}
\includegraphics[width=0.5\textwidth]{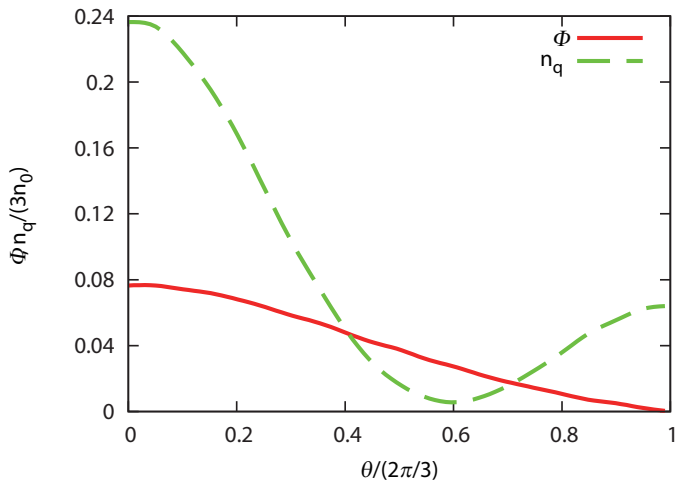}
\end{center}
\caption{$\theta$ dependence of $\Phi$ and the quark number density
 $n_q$ at $(\mu , T)=(260~{\rm MeV},100~{\rm MeV})$ in the PNJL model.
The solid line stands for $\Phi$, while the dashed line corresponds to
 $n_q$. The $n_q$ is divided by $3n_0$, where $n_0$ is the normal
 nuclear matter density.
}
\label{Phi_nq_1}
\end{figure}

Figure~\ref{Phi_nq_1} shows $\theta$ dependence of $\Phi$ and
the quark number density $n_q$
at the point $(\mu , T)=(260~{\rm MeV},100~{\rm MeV})$.
At $\theta =2\pi /3$, this point is in the perfectly confined phase with
chiral symmetry breaking.
We see that the $\theta$-dependence is smooth in a range of
$\theta =0 \sim 2\pi /3$.
Hence, the approach mentioned above may be valid for this point because there is no singularity. 
In particular, $\Phi$ increases monotonically as $\theta$ decreases from
$\theta=2\pi/3$. 
This may suggest that only the lower derivative terms are
needed in the Taylor expansion (\ref{Taylor}) to evaluate $\Phi$
at smaller $\theta$.
On the contrary, $n_q$ has a minimum at $\theta \sim 1.3$.
In the Stefan-Boltzmann limit, $n_q$ is given by
\begin{eqnarray}
n_q&=&(\mu +i\theta T)T^2+(\mu+i\theta T)^3
\nonumber\\
&&+(\mu -i\theta T)T^2+(\mu -i\theta T)^3
\nonumber\\
&=&2\left( \mu T^2+\mu^3-3\mu \theta^2 T^2\right),
\label{nq}
\end{eqnarray}
and it decreases monotonically as $\theta$ increases from $\theta=0$.
As easily seen in Eq. (\ref{det_h_q}), however, the $\theta$ term is
suppressed by strong confinement near $\theta=2\pi/3$. 
As a consequence of the suppression, $n_q$ is considered to have a minimum at $\theta
\sim 1.3$.
The fact that $\theta$ dependence is not monotonic for $n_q$ means that
the higher derivative terms are necessary in (\ref{Taylor})
to evaluate $n_q$ at smaller $\theta$.

\begin{figure}[htbp]
\begin{center}
\includegraphics[width=0.5\textwidth]{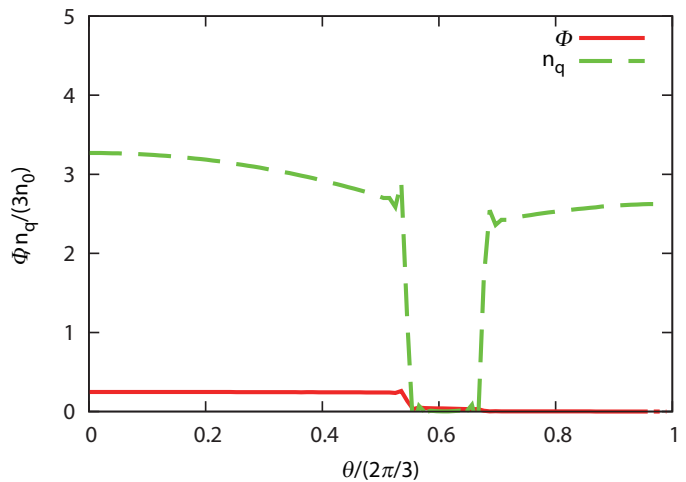}
\end{center}
\caption{$\theta$ dependence of $\Phi$ and 
the quark number density $n_q$ at 
$(\mu ,T) =(300~{\rm MeV},100~{\rm MeV})$ in the PNJL model. 
See Fig. \ref{Phi_nq_1} for the definition of lines.
}
\label{Phi_nq_2}
\end{figure}

Figure~\ref{Phi_nq_2} is the same as Fig.~\ref{Phi_nq_1} but at the
point $(\mu ,T) =(300~{\rm MeV},100~{\rm MeV})$.
At $\theta =2\pi /3$, this point is in the perfectly confined phase with
chiral symmetry restoration.
We see that both $\Phi$ and $n_q$ have discontinuities as a function of
$\theta$.
Hence, the Taylor-expansion approach of (\ref{Taylor}) does not work at
this point.
Thus, the Taylor-expansion approach of (\ref{Taylor}) may work, if the
system is in the perfectly confined phase with chiral symmetry breaking
in the case of $Z_3$-QCD.

We remark that, in the numerical calculations for Figs.~\ref{Phi_nq_1}
and \ref{Phi_nq_2}, we assumed that the diquark condensate does not
appear in the whole region of $\theta =0\sim 2\pi/3$. 
Since the analytic form of the quark spectrum $E_j$ is not known, we
must use numerical differentiation to determine the quark number density
$n_q$, when $\Delta_l \neq 0$. 
We found that it is very difficult to perform such differentiation with
high accuracy in the low temperature region where $T \leq 100$MeV.
Then, we adopt the assumption mentioned above and performed the
approximated calculations. 
This assumption is very natural for $\mu =260$MeV where the chiral
symmetry is broken due to large quark mass $M$ but may not be valid for
$\mu =300$MeV where $M$ is small. 
If the diquark condensate appears at some value of $\theta$ in the case
of $(\mu ,T) =(300{\rm MeV}, 100{\rm MeV})$, $n_q$ and $\Phi$ is not
analytic there. 
Hence, the conclusion that the Taylor-expansion approach of
(\ref{Taylor}) does not work at $(\mu ,T) =(300{\rm MeV}, 100{\rm MeV})$
is not changed. 

\section{Summary}
\label{Summary}

In summary, we investigate QCD at large $\mu/T$ by using the $SU(3)$
gauge theory with the flavor-dependent twist boundary condition
(FTBC). 
The theory agrees with QCD at $\theta=0$ and becomes $Z_3$
symmetric at $\theta=2\pi/3$. 
In $Z_3$-QCD, one can make clear discussion on the relation
between $Z_3$-symmetry and diquark and chiral condensates.

In $Z_3$-QCD, there can exist a perfectly confined phase
where $\Phi=0$.
Since the diquark condensate $\Delta$ is an order parameter
for both $Z_3$ symmetry and CSC,
$\Delta$ is finite (zero) when $\Phi$ is finite (zero).
The perfectly confined phase with $\Phi=0$ thus
never coexists with the CSC phase with $\Delta \neq 0$. 
Meanwhile, the chiral condensate as an order parameter for the chiral
transition is $Z_3$ invariant. Hence the chiral transition can take
place in the perfectly confined phase.

The phase diagram  was numerically investigated for $Z_3$-QCD with the
PNJL model.
The perfectly confined phase with $\Phi=0$ and $\Delta=0$ appears at $T
\lsim 200$~MeV, and the chiral restoration occurs at $\mu \approx
300$~MeV in the phase.
Hence, the perfectly confined phase is divided into the perfectly
confined phase without chiral symmetry restoration in
a region of $\mu \lsim 300$~MeV and $T \lsim 200$~MeV and
the perfectly confined phase with chiral symmetry restoration
in a region of $\mu \gsim 300$~MeV and $100 \lsim T \lsim 200$~MeV.
In a region of $\mu \gsim 300$~MeV and $T \lsim 100$~MeV,
CSC phases with finite $\Delta$ and small $\Phi$ appear, since they
are more stable than the perfectly confined phase with chiral symmetry
restoration.
This basic structure remains in QCD, although the phase structure is
much more complicated in $Z_3$-QCD than that in QCD.
The phase diagram for QCD is thus a remnant of that for $Z_3$-QCD.

We also showed that in the perfectly confined phase the sign problem may
become less serious because of $\Phi=0$, using the heavy quark model.
This may make LQCD simulations feasible in the perfectly confined phase.
Particularly when the system is in the perfectly confined phase without
chiral symmetry restoration, the system changes continuously
from $Z_3$-QCD to QCD by varying $\theta$ from $2\pi/3$ to 0.
Using this property, we proposed the Taylor-expansion method
of (\ref{Taylor}) to estimate observables for QCD from those for
$Z_3$-QCD.
In the perfectly confined phase with chiral symmetry restoration,
however, the system does not change smoothly as $\theta$ varies from
$2\pi/3$ to 0. Therefore, the lattice QCD framework proposed above may
not work. Further study is necessary for this region.

In this paper, we considered three degenerate flavor QCD
as the first step. Even in the 2+1 flavor case, there exists
a point of $\Phi =0$ in $\mu$--$T$ plane for some $\theta$ near
$2\pi/3$~\cite{Kouno_TBC_2}.
For example, in the case of infinite $m_s$ (the two-flavor case),
such a point appears at $\theta \sim {\pi/{2}}$.
Hence, we can expect that the present Taylor expansion method is
applicable for realistic 2+1 flavor QCD.

\noindent
\begin{acknowledgments}
The authors are thankful especially to I.-O. Stamatescu for useful
 information  on their analyses based on the heavy quark model for
 finite density
and to Y. Sakai and T. Sasaki for many fruitful discussions on the FTBC
 model.
They also thank A. Nakamura, Y. Taniguchi, T. Saito, E.Ito, K. Nagata,
 R. Fukuda and A. Suzuki for variable comments.
H.K. also thanks M. Imachi, H. Yoneyama, H. Aoki, M. Tachibana and
 T. Makiyama for useful discussions.
Four of the authors (H. K., K. K., J. T. and M. Y.) are supported
by Grant-in-Aid for Scientific Research (No.26400279, No. 26-1717,
 No. 25-3944 and No.26400278) from Japan Society for the Promotion of
 Science (JSPS).
The numerical calculations were partially performed by using SX-ACE at
 CMC, Osaka University.
\end{acknowledgments}



\begin{thebibliography}{19}
\expandafter\ifx\csname natexlab\endcsname\relax\def\natexlab#1{#1}\fi
\expandafter\ifx\csname bibnamefont\endcsname\relax
  \def\bibnamefont#1{#1}\fi
\expandafter\ifx\csname bibfnamefont\endcsname\relax
  \def\bibfnamefont#1{#1}\fi
\expandafter\ifx\csname citenamefont\endcsname\relax
  \def\citenamefont#1{#1}\fi
\expandafter\ifx\csname url\endcsname\relax
  \def\url#1{\texttt{#1}}\fi
\expandafter\ifx\csname urlprefix\endcsname\relax\def\urlprefix{URL }\fi
\providecommand{\bibinfo}[2]{#2}
\providecommand{\eprint}[2][]{\url{#2}}
%
\bibitem[{\citenamefont{Fodor}(2002)}]{Fodor}
\bibinfo{author}{\bibfnamefont{Z.}~\bibnamefont{Fodor}}, 
\bibnamefont{and} 
\bibinfo{author}{\bibfnamefont{S.}~\bibnamefont{D.}~\bibnamefont{Katz}},  
\bibinfo{journal}{Phys. Lett.\ B} \textbf{\bibinfo{volume}{534}}, 
\bibinfo{pages}{87} (\bibinfo{year}{2002}). 
%
\bibitem[{\citenamefont{Allton}(2004)}]{Allton}
\bibinfo{author}{\bibfnamefont{C.}~\bibfnamefont{R.}~\bibnamefont{Allton}},
\bibinfo{author}{\bibfnamefont{S.}~\bibnamefont{Ejiri}},
\bibinfo{author}{\bibfnamefont{S.}~\bibfnamefont{J.}~\bibnamefont{Hands}},
\bibinfo{author}{\bibfnamefont{O.}~\bibnamefont{Kaczmarek}},
\bibinfo{author}{\bibfnamefont{F.}~\bibnamefont{Karsch}},
\bibinfo{author}{\bibfnamefont{E.}~\bibnamefont{Laermann}},
\bibinfo{author}{\bibfnamefont{Ch.}~\bibnamefont{Schmidt}},
\bibnamefont{and} 
\bibinfo{author}{\bibfnamefont{L.}~\bibnamefont{Scorzato}},
  \bibinfo{journal}{Phys. Rev. D} \textbf{\bibinfo{volume}{66}},
  \bibinfo{pages}{074507} (\bibinfo{year}{2002}). 

%
\bibitem[{\citenamefont{Ejiri et al.}(2004)}]{Ejiri_density}
\bibinfo{author}{\bibfnamefont{S.}~\bibnamefont{Ejiri et al.}}, 
\bibinfo{journal}{Phys. Rev. D} \textbf{\bibinfo{volume}{82}},
\bibinfo{pages}{014508} (\bibinfo{year}{2010}). 
%
\bibitem[{\citenamefont{Forcrand and Philipsen}(2002)}]{FP}
\bibinfo{author}{\bibfnamefont{P.}~\bibnamefont{de}~\bibnamefont{Forcrand}} 
\bibnamefont{and}
\bibinfo{author}{\bibfnamefont{O.}~\bibnamefont{Philipsen}},  
\bibinfo{journal}{Nucl. Phys. } \textbf{\bibinfo{volume}{B642}},
\bibinfo{pages}{290} (\bibinfo{year}{2002}). 
%
\bibitem[{\citenamefont{Elia and Lombardo}(2003)}]{D'Elia}
\bibinfo{author}{\bibfnamefont{M.}~\bibnamefont{D'Elia}} \bibnamefont{and}
\bibinfo{author}{\bibfnamefont{M.}~\bibfnamefont{P.}~\bibnamefont{Lombardo}},  
\bibinfo{journal}{Phys. Rev.\  D} \textbf{\bibinfo{volume}{67}},
\bibinfo{pages}{014505} (\bibinfo{year}{2003}). 
%
\bibitem[{\citenamefont{D'Elia et al}(2009)}]{D'Elia3}
\bibinfo{author}{\bibfnamefont{M.}~\bibnamefont{D'Elia}} \bibnamefont{and}
\bibinfo{author}{\bibfnamefont{F.}~\bibnamefont{Sanfilippo}},
\bibinfo{journal}{Phys.\ Rev.\ D} \textbf{\bibinfo{volume}{80}},
\bibinfo{pages}{111501} (\bibinfo{year}{2009}).
%
\bibitem[{\citenamefont{FP2010}(2009)}]{FP2010}
\bibinfo{author}{\bibfnamefont{P.}~\bibnamefont{de}~\bibnamefont
{Forcrand}} 
\bibnamefont{and}
\bibinfo{author}{\bibfnamefont{O.}~\bibnamefont{Philipsen}},  
\bibinfo{journal}{Phys.\ Rev.\ Lett. } \textbf{\bibinfo{volume}{105}},
\bibinfo{pages}{152001} (\bibinfo{year}{2010}). 
%
\bibitem{Nagata}
\bibinfo{author}{\bibfnamefont{K.}~\bibnamefont{Nagata}}
\bibnamefont{and}
\bibinfo{author}{\bibfnamefont{A.}~\bibnamefont{Nakamura}},  
\bibinfo{journal}{Phys. Rev. D} \textbf{\bibinfo{volume}{83}},
\bibinfo{pages}{114507} (\bibinfo{year}{2011}). 
%
\bibitem{Takahashi}
\bibinfo{author}{\bibfnamefont{J.}~\bibnamefont{Takahashi}}, 
\bibinfo{author}{\bibfnamefont{K.}~\bibnamefont{Nagata}}, 
\bibinfo{author}{\bibfnamefont{T.}~\bibnamefont{Saito}},
\bibinfo{author}{\bibfnamefont{A.}~\bibnamefont{Nakamura}}, 
\bibinfo{author}{\bibfnamefont{T.}~\bibnamefont{Sasaki}}, 
\bibinfo{author}{\bibfnamefont{H.}~\bibnamefont{Kouno}}, 
\bibnamefont{and} 
\bibinfo{author}{\bibfnamefont{M.}~\bibnamefont{Yahiro}} 
\bibinfo{journal}{Phys. Rev. D} \textbf{\bibinfo{volume}{88}},
\bibinfo{pages}{114504} (\bibinfo{year}{2013}); 
\bibinfo{author}{\bibfnamefont{J.}~\bibnamefont{Takahashi}}, 
\bibinfo{author}{\bibfnamefont{H.}~\bibnamefont{Kouno}}, 
\bibnamefont{and} 
\bibinfo{author}{\bibfnamefont{M.}~\bibnamefont{Yahiro}} 
\bibinfo{journal}{Phys. Rev. D} \textbf{\bibinfo{volume}{91}},
\bibinfo{pages}{014501} (\bibinfo{year}{2015}).
%
\bibitem{Aarts_CLE_1}
G. Aarts, 
\bibinfo{journal}{Phys.\ Rev.\ Lett. } \textbf{\bibinfo{volume}{102}},
\bibinfo{pages}{131601} (\bibinfo{year}{2009}). 
%
\bibitem{Aarts_CLE_2} 
G. Aarts, L. Bongiovanni, E. Seiler, D. Sexty, and I.-O. Stamatescu, 
\bibinfo{journal}{Eur.\ Phys.\ J.\ A } \textbf{\bibinfo{volume}{49}},
\bibinfo{pages}{89} (\bibinfo{year}{2013}). 
%
\bibitem{Sexty} 
D. Sexty,  
\bibinfo{journal}{Phys. Lett.\ B} \textbf{\bibinfo{volume}{729}},
\bibinfo{pages}{108} (\bibinfo{year}{2014}). 
%
\bibitem[{\citenamefont{Greensite}(2014)}]{Greensite}
\bibinfo{author}{\bibfnamefont{J.}~\bibnamefont{Greensite}}, 
arXiv:1406.4558 [hep-lat] (2014).  
%
\bibitem{Aarts_CLE_3} 
G. Aarts, F. Attanasio, B. J\"{a}ger, E. Seiler, D. Sexty, 
and I.-O. Stamatescu, arXiv:1411.2632 [hep-lat](2014).
%
\bibitem{Aurora_thimbles} 
M. Cristoforetti et al., 
\bibinfo{journal}{Phys.\ Rev.\ D } \textbf{\bibinfo{volume}{86}},
\bibinfo{pages}{074506} (\bibinfo{year}{2012}). 
%
\bibitem{Fujii_thimbles} 
H. Fujii, D. Honda, M. Kato, Y. Kikukawa, S. Komatsu and T. Sano, 
\bibinfo{journal}{JHEP} \textbf{\bibinfo{volume}{1310}},
\bibinfo{pages}{147} (\bibinfo{year}{2013}). 
%
\bibitem[{\citenamefont{Meisinger et al.}(1996)}]{Meisinger}
\bibinfo{author}{\bibfnamefont{P.}~\bibnamefont{N.}}~\bibnamefont{Meisinger},
\bibnamefont{and}
\bibinfo{author}{\bibfnamefont{M.}~\bibnamefont{C.}}~\bibnamefont{Ogilvie},  
  \bibinfo{journal}{Phys. Lett.\ B} \textbf{\bibinfo{volume}{379}},
  \bibinfo{pages}{163} (\bibinfo{year}{1996}). 
%
\bibitem[{\citenamefont{Dumitru}(2002)}]{Dumitru}
\bibinfo{author}{\bibfnamefont{A.}~\bibnamefont{Dumitru}},
\bibnamefont{and}
\bibinfo{author}{\bibfnamefont{R.}~\bibfnamefont{D.}~\bibnamefont{Pisarski}},  
\bibinfo{journal}{Phys.\ Rev.\  D} \textbf{\bibinfo{volume}{66}},
\bibinfo{pages}{096003} (\bibinfo{year}{2002}); 
\bibinfo{author}{\bibfnamefont{A.}~\bibnamefont{Dumitru}},
\bibinfo{author}{\bibfnamefont{Y.}~\bibnamefont{Hatta}},
\bibinfo{author}{\bibfnamefont{J.}~\bibnamefont{Lenaghan}},
\bibinfo{author}{\bibfnamefont{K.}~\bibnamefont{Orginos}},
\bibnamefont{and}
\bibinfo{author}{\bibfnamefont{R.}~\bibfnamefont{D.}~\bibnamefont{Pisarski}},  
\bibinfo{journal}{Phys.\ Rev.\  D} \textbf{\bibinfo{volume}{70}},
\bibinfo{pages}{034511} (\bibinfo{year}{2004}); 
\bibinfo{author}{\bibfnamefont{A.}~\bibnamefont{Dumitru}},
\bibinfo{author}{\bibfnamefont{R.}~\bibfnamefont{D.}~\bibnamefont{Pisarski}},  
\bibnamefont{and}
\bibinfo{author}{\bibfnamefont{D.}~\bibnamefont{Zschiesche}},  
\bibinfo{journal}{Phys.\ Rev.\  D} \textbf{\bibinfo{volume}{72}},
\bibinfo{pages}{065008} (\bibinfo{year}{2005}).
%
\bibitem[{\citenamefont{Fukushima}(2004)}]{Fukushima}
\bibinfo{author}{\bibfnamefont{K.}~\bibnamefont{Fukushima}}, 
  \bibinfo{journal}{Phys. Lett.\ B} \textbf{\bibinfo{volume}{591}},
  \bibinfo{pages}{277} (\bibinfo{year}{2004}).. 
%
\bibitem[{\citenamefont{Ratti et al.}(2006)}]{Ratti}
\bibinfo{author}{\bibfnamefont{C.}~\bibnamefont{Ratti}},
\bibinfo{author}{\bibfnamefont{M.}~\bibfnamefont{A.}~\bibnamefont{Thaler}},
\bibnamefont{and}
\bibinfo{author}{\bibfnamefont{W.}~\bibnamefont{Weise}},  
  \bibinfo{journal}{Phys. Rev.\ D} \textbf{\bibinfo{volume}{73}},
  \bibinfo{pages}{014019} (\bibinfo{year}{2006}); 
\bibinfo{author}{\bibfnamefont{C.}~\bibnamefont{Ratti}},
\bibinfo{author}{\bibfnamefont{S.}~\bibnamefont{R\"{o}{\ss}ner}},
\bibinfo{author}{\bibfnamefont{M.}~\bibfnamefont{A.}~\bibnamefont{Thaler}},
\bibnamefont{and}
\bibinfo{author}{\bibfnamefont{W.}~\bibnamefont{Weise}},  
  \bibinfo{journal}{Eur. Phys. J.\ C} \textbf{\bibinfo{volume}{49}},
  \bibinfo{pages}{213} (\bibinfo{year}{2007}). 
%
\bibitem[{\citenamefont{Megias al.}(2006)}]{Megias}
\bibinfo{author}{\bibfnamefont{E.}~\bibnamefont{Megias}}, 
\bibinfo{author}{\bibfnamefont{E.}~\bibnamefont{Ruiz Arriola}}, 
\bibnamefont{and}
\bibinfo{author}{\bibfnamefont{L.~L.}~\bibnamefont{ Salcedo}},
\bibinfo{journal}{Phys.\ Rev.\ D} \textbf{\bibinfo{volume}{74}},
\bibinfo{pages}{065005} (\bibinfo{year}{2006}).
%
\bibitem[{\citenamefont{Rossner et al.}(2007)}]{Rossner}
\bibinfo{author}{\bibfnamefont{S.}~\bibnamefont{R\"{o}{\ss}ner}},
\bibinfo{author}{\bibfnamefont{C.}~\bibnamefont{Ratti}},
\bibnamefont{and}
\bibinfo{author}{\bibfnamefont{W.}~\bibnamefont{Weise}},  
  \bibinfo{journal}{Phys. Rev.\ D} \textbf{\bibinfo{volume}{75}},
  \bibinfo{pages}{034007} (\bibinfo{year}{2007}). 
%
\bibitem[{\citenamefont{Schaefer}(2007)}]{Schaefer}
\bibinfo{author}{\bibfnamefont{B.}~\bibfnamefont{-J.}~\bibnamefont{Schaefer}},
\bibinfo{author}{\bibfnamefont{J.}~\bibfnamefont{M.}~\bibnamefont{Pawlowski}},
\bibnamefont{and}
\bibinfo{author}{\bibfnamefont{J.}~\bibnamefont{Wambach}},  
  \bibinfo{journal}{Phys.\ Rev.\  D} \textbf{\bibinfo{volume}{76}},
  \bibinfo{pages}{074023} (\bibinfo{year}{2007}).
%
\bibitem[{\citenamefont{Abuki et al}(2008)}]{Abuki}
\bibinfo{author}{\bibfnamefont{H.}~\bibnamefont{Abuki}},
\bibinfo{author}{\bibfnamefont{R.}~\bibnamefont{Anglani}},
\bibinfo{author}{\bibfnamefont{R.}~\bibnamefont{Gatto}},
\bibinfo{author}{\bibfnamefont{G.}~\bibnamefont{Nardulli}},
\bibnamefont{and}
\bibinfo{author}{\bibfnamefont{M.}~\bibnamefont{Ruggieri}},
\bibinfo{journal}{Phys.\ Rev.\  D} \textbf{\bibinfo{volume}{78}},
\bibinfo{pages}{034034} (\bibinfo{year}{2008}). 
%
\bibitem[{\citenamefont{Fukushima}(2004)}]{Fukushima2}
\bibinfo{author}{\bibfnamefont{K.}~\bibnamefont{Fukushima}}, 
  \bibinfo{journal}{Phys.\ Rev.\  D} \textbf{\bibinfo{volume}{77}},
  \bibinfo{pages}{114028} (\bibinfo{year}{2008}).
%
\bibitem[{\citenamefont{Kashiwa et al}(2008)}]{Kashiwa1}
\bibinfo{author}{\bibfnamefont{K.}~\bibnamefont{Kashiwa}}, 
\bibinfo{author}{\bibfnamefont{H.}~\bibnamefont{Kouno}}, 
\bibinfo{author}{\bibfnamefont{M.}~\bibnamefont{Matsuzaki}}, 
\bibnamefont{and}
\bibinfo{author}{\bibfnamefont{M.}~\bibnamefont{Yahiro}},
  \bibinfo{journal}{Phys.\ Lett.\ B} \textbf{\bibinfo{volume}{662}},
  \bibinfo{pages}{26} (\bibinfo{year}{2008}).
%
\bibitem[{\citenamefont{Sakai et al}(2008)}]{Sakai}
\bibinfo{author}{\bibfnamefont{Y.}~\bibnamefont{Sakai}},
\bibinfo{author}{\bibfnamefont{K.}~\bibnamefont{Kashiwa}}, 
\bibinfo{author}{\bibfnamefont{H.}~\bibnamefont{Kouno}}, 
\bibnamefont{and}
\bibinfo{author}{\bibfnamefont{M.}~\bibnamefont{Yahiro}},
\bibinfo{journal}{Phys.\ Rev.\  D} \textbf{\bibinfo{volume}{77}},
\bibinfo{pages}{051901} (\bibinfo{year}{2008});
\bibinfo{journal}{Phys.\ Rev.\  D} \textbf{\bibinfo{volume}{78}},
\bibinfo{pages}{ 036001} (\bibinfo{year}{2008}). 
%
\bibitem{McLerran_largeNc}
\bibinfo{author}{\bibfnamefont{L.}~\bibnamefont{McLerran}}
\bibinfo{author}{\bibfnamefont{K.}~\bibnamefont{Redlich}}
\bibnamefont{and}
\bibinfo{author}{\bibfnamefont{C.}~\bibnamefont{Sasaki}}, 
  \bibinfo{journal}{Nucl. Phys.\  A} \textbf{\bibinfo{volume}{824}},
  \bibinfo{pages}{86} (\bibinfo{year}{2009}).
%
\bibitem[{\citenamefont{Sasaki et al.}(2011)}]{Sasaki-T_Nf3}
\bibinfo{author}{\bibfnamefont{T.}~\bibnamefont{Sasaki}}, 
\bibinfo{author}{\bibfnamefont{Y.}~\bibnamefont{Sakai}}, 
\bibinfo{author}{\bibfnamefont{H.}~\bibnamefont{Kouno}}, 
\bibnamefont{and}
\bibinfo{author}{\bibfnamefont{M.}~\bibnamefont{Yahiro}}, 
\bibinfo{journal}{Phys. Rev. \ D} 
\textbf{\bibinfo{volume}{84}},
\bibinfo{pages}{091901} (\bibinfo{year}{2011}). 
%
\bibitem[{\citenamefont{Polyakov}(1978)}]{Polyakov}
\bibinfo{author}{\bibfnamefont{A.}~\bibfnamefont{M.}~\bibnamefont{Polyakov}}, 
\bibinfo{journal}{Phys. Lett.} \textbf{\bibinfo{volume}{72B}},
\bibinfo{pages}{477} (\bibinfo{year}{1978}).
%
%
\bibitem{YAoki_nature}
Y. Aoki, G. Endr\"{o}di, Z. Fodor, S. D. Katz and K. K. Szab\'{o}, 
Nature {\bf 443}, 675 (2006). 
%


\bibitem[{\citenamefont{Kouno et al}(2012)}]{Kouno_TBC}
\bibinfo{author}{\bibfnamefont{H.}~\bibnamefont{Kouno}}, 
\bibinfo{author}{\bibfnamefont{Y.}~\bibnamefont{Sakai}}, 
\bibinfo{author}{\bibfnamefont{T.}~\bibnamefont{Makiyama}},
\bibinfo{author}{\bibfnamefont{K.}~\bibnamefont{Tokunaga}},
\bibinfo{author}{\bibfnamefont{T.}~\bibnamefont{Sasaki}},   
\bibnamefont{and}
\bibinfo{author}{\bibfnamefont{M.}~\bibnamefont{Yahiro}},  
\bibinfo{journal}{J. Phys. G: Nucl. Part. Phys.} \textbf{\bibinfo{volume}{39}},
\bibinfo{pages}{085010} (\bibinfo{year}{2012}). 
%
\bibitem[{\citenamefont{Kouno et al}(2012)}]{Sakai_TBC}
\bibinfo{author}{\bibfnamefont{Y.}~\bibnamefont{Sakai}}, 
\bibinfo{author}{\bibfnamefont{H.}~\bibnamefont{Kouno}}, 
\bibinfo{author}{\bibfnamefont{T.}~\bibnamefont{Sasaki}},   
\bibnamefont{and}
\bibinfo{author}{\bibfnamefont{M.}~\bibnamefont{Yahiro}},  
\bibinfo{journal}{Phys. Lett.\ B} \textbf{\bibinfo{volume}{718}},
\bibinfo{pages}{130} (\bibinfo{year}{2012}). 
%
\bibitem[{\citenamefont{Kouno et al}(2013)}]{Kouno_adjoint}
\bibinfo{author}{\bibfnamefont{H.}~\bibnamefont{Kouno}}, 
\bibinfo{author}{\bibfnamefont{K.}~\bibnamefont{Kashiwa}}, 
\bibinfo{author}{\bibfnamefont{T.}~\bibnamefont{Misumi}},
\bibinfo{author}{\bibfnamefont{T.}~\bibnamefont{Makiyama}},
\bibinfo{author}{\bibfnamefont{T.}~\bibnamefont{Sasaki}},   
\bibnamefont{and}
\bibinfo{author}{\bibfnamefont{M.}~\bibnamefont{Yahiro}}, 
\bibinfo{journal}{Phys. Rev.\ D} \textbf{\bibinfo{volume}{88}},
\bibinfo{pages}{016002} (\bibinfo{year}{2013}). 
%
\bibitem[{\citenamefont{Kouno et al}(2013)}]{Kouno_TBC_2}
\bibinfo{author}{\bibfnamefont{H.}~\bibnamefont{Kouno}}, 
\bibinfo{author}{\bibfnamefont{T.}~\bibnamefont{Makiyama}},
\bibinfo{author}{\bibfnamefont{T.}~\bibnamefont{Sasaki}}, 
\bibinfo{author}{\bibfnamefont{Y.}~\bibnamefont{Sakai}},   
\bibnamefont{and}
\bibinfo{author}{\bibfnamefont{M.}~\bibnamefont{Yahiro}}, 
\bibinfo{journal}{J. Phys. G: Nucl. Part. Phys.} \textbf{\bibinfo{volume}{40}},
\bibinfo{pages}{095003} (\bibinfo{year}{2013}). 
%
%
\bibitem{BW_C_1}
S. Borsanyi, Z. Fodor, C. Hoelbling, S.D. Katz, S. Krieg, C. Ratti, K. K. Szabo, 
JHEP {\bf 1009:073} (2010). 
%
\bibitem{HQCD_1}
A. Bazavov et al., Phys. Rev. D{\bf 85}, 054503 (2012). 
JHEP {\bf 1009:073} (2010). 
%
\bibitem[{\citenamefont{Roberge and Weiss}(1986)}]{RW}
\bibinfo{author}{\bibfnamefont{A.}~\bibnamefont{Roberge}} \bibnamefont{and}
\bibinfo{author}{\bibfnamefont{N.}~\bibnamefont{Weiss}},  
\bibinfo{journal}{Nucl. Phys. } \textbf{\bibinfo{volume}{B275}},
\bibinfo{pages}{734} (\bibinfo{year}{1986}). 














%
\bibitem[{\citenamefont{Boyd et al.}(1996)}]{Boyd}
\bibinfo{author}{\bibfnamefont{G.}~\bibnamefont{Boyd}},
\bibinfo{author}{\bibfnamefont{J.}~\bibnamefont{Engels}},
\bibinfo{author}{\bibfnamefont{F.}~\bibnamefont{Karsch}},
\bibinfo{author}{\bibfnamefont{E.}~\bibnamefont{Laermann}},
\bibinfo{author}{\bibfnamefont{C.}~\bibnamefont{Legeland}},
\bibinfo{author}{\bibfnamefont{M.}~\bibnamefont{L\"{u}tgemeier}},
\bibnamefont{and}
\bibinfo{author}{\bibfnamefont{B.}~\bibnamefont{Petersson}},
 \bibinfo{journal}{Nucl. Phys.} \textbf{\bibinfo{volume}{B469}},
\bibinfo{pages}{419} (\bibinfo{year}{1996}). 
%
\bibitem[{\citenamefont{Kaczmarek}(2002)}]{Kaczmarek}
\bibinfo{author}{\bibfnamefont{O.}~\bibnamefont{Kaczmarek}},
\bibinfo{author}{\bibfnamefont{F.}~\bibnamefont{Karsch}},
\bibinfo{author}{\bibfnamefont{P.}~\bibnamefont{Petreczky}},
\bibnamefont{and}
\bibinfo{author}{\bibfnamefont{F.}~\bibnamefont{Zantow}},  
  \bibinfo{journal}{Phys. Lett.\ B} \textbf{\bibinfo{volume}{543}},
  \bibinfo{pages}{41} (\bibinfo{year}{2002}).
%
\bibitem[{\citenamefont{Borsanyi etal}(2010)}]{Borsanyi}
\bibinfo{author}{\bibfnamefont{S.}~\bibnamefont{Bors\'{a}nyi}}, 
\bibinfo{author}{\bibfnamefont{Z.}~\bibnamefont{Fodor}}, 
\bibinfo{author}{\bibfnamefont{C.}~\bibnamefont{Hoelbling}}, 
\bibinfo{author}{\bibfnamefont{S.}~\bibnamefont{D.}~\bibnamefont{Katz}}, 
\bibinfo{author}{\bibfnamefont{S.}~\bibnamefont{Krieg}}, 
\bibinfo{author}{\bibfnamefont{C.}~\bibnamefont{Ratti}}, 
\bibnamefont{and} 
\bibinfo{author}{\bibfnamefont{K.}~\bibnamefont{K.}~\bibnamefont{Szabo}},  
\bibinfo{howpublished}{arXiv:1005.3508 [hep-lat]} (\bibinfo{year}{2010}). 
%
\bibitem[{\citenamefont{Soeldner}(2010)}]{Soeldner}
\bibinfo{author}{\bibfnamefont{W.}~\bibnamefont{S\"{o}ldner}}, 
\bibinfo{howpublished}{arXiv:1012.4484 [hep-lat]} (\bibinfo{year}{2010}). 
%
\bibitem[{\citenamefont{Kanaya}(2010)}]{Kanaya}
\bibinfo{author}{\bibfnamefont{K.}~\bibnamefont{Kanaya}}, 
\bibinfo{howpublished}{arXiv:hep-ph/1012.4235 [hep-ph]} (\bibinfo{year}{2010});
\bibinfo{howpublished}{arXiv:hep-ph/1012.4247 [hep-lat]} (\bibinfo{year}{2010}).
%
\bibitem[{\citenamefont{Ruster et al}(2008)}]{Ruster}
\bibinfo{author}{\bibfnamefont{S.~B.}~\bibnamefont{R\"{u}ster}},
\bibinfo{author}{\bibfnamefont{V.}~\bibnamefont{Werth}},
\bibinfo{author}{\bibfnamefont{M.}~\bibnamefont{Buballa}},
\bibinfo{author}{\bibfnamefont{I.~A.}~\bibnamefont{Shovkovy}},
\bibnamefont{and}
\bibinfo{author}{\bibfnamefont{D.~H.}~\bibnamefont{Rischke}},
\bibinfo{journal}{Phys.\ Rev.\  D} \textbf{\bibinfo{volume}{72}},
\bibinfo{pages}{034004} (\bibinfo{year}{2005}). 
%
\bibitem[{\citenamefont{Kobayashi and Maskawa}(1970)}]{KMK}
\bibinfo{author}{\bibfnamefont{M.}~\bibnamefont{Kobayashi}}, 
\bibnamefont{and}
\bibinfo{author}{\bibfnamefont{T.}~\bibnamefont{Maskawa}},
  \bibinfo{journal}{Prog. Theor. Phys. } \textbf{\bibinfo{volume}{44}},
  \bibinfo{pages}{1422} (\bibinfo{year}{1970});
\bibinfo{author}{\bibfnamefont{M.}~\bibnamefont{Kobayashi}}, 
\bibinfo{author}{\bibfnamefont{H.}~\bibnamefont{Kondo}},
\bibnamefont{and}
\bibinfo{author}{\bibfnamefont{T.}~\bibnamefont{Maskawa}},
  \bibinfo{journal}{Prog. Theor. Phys. } \textbf{\bibinfo{volume}{45}},
  \bibinfo{pages}{1955} (\bibinfo{year}{1971}). 
%
\bibitem[{\citenamefont{'t Hooft}(1976)}]{tHooft}
\bibinfo{author}{\bibfnamefont{G.}~\bibnamefont{'t Hooft}},
  \bibinfo{journal}{Phys. Rev.\ Lett.} \textbf{\bibinfo{volume}{37}},
  \bibinfo{pages}{8} (\bibinfo{year}{1976});
  \bibinfo{journal}{Phys. Rev.\ D} \textbf{\bibinfo{volume}{14}},
  \bibinfo{pages}{3432} (\bibinfo{year}{1976});
  \textbf{\bibinfo{volume}{18}},
  \bibinfo{pages}{2199(E)} (\bibinfo{year}{1978}).
%
\bibitem{Rehberg}
\bibinfo{author}{\bibfnamefont{P.}~\bibnamefont{Rehberg}},
\bibinfo{author}{\bibfnamefont{S.P.}~\bibnamefont{Klevansky}}
\bibnamefont{and}
\bibinfo{author}{\bibfnamefont{J.}~\bibnamefont{H\"{u}fner}}, 
  \bibinfo{journal}{Phys.\ Rev.\  C} \textbf{\bibinfo{volume}{53}},
  \bibinfo{pages}{410} (\bibinfo{year}{1996}). 














\end{thebibliography}
\end{document}